\documentclass[fleqn,12pt,twoside]{article}
\usepackage{espcrc1}

\usepackage{graphicx}
\usepackage[figuresright]{rotating}

\newcommand{\AmS}{{\protect\the\textfont2
  A\kern-.1667em\lower.5ex\hbox{M}\kern-.125emS}}

\def\et{$\it{et~al.}$}
\def\pik{($\pi^{+}, K^{+}$)}

\def\pim{$\pi^{-}$}

\def\piz{$\pi^0$}

\def\HeLam{$^{5}_{\Lambda}$He}
\def\LiLam{$^{6}_{\Lambda}$Li}
\def\CLam{$^{12}_{\Lambda}$C}

\def\GLam{$\Gamma_{\Lambda}$}
\def\Gtot{$\Gamma_{tot}$}
\def\Gpiz{$\Gamma_{\pi^{0}}$}
\def\Gpim{$\Gamma_{\pi^{-}}$}
\def\Gn{$\Gamma_{n}$}
\def\Gp{$\Gamma_{p}$}

\def\Gnm{$\Gamma_{nm}$}

\title{\piz ~decay branching ratios
of \HeLam ~and \CLam ~hypernuclei}
\author
{ S.~Okada$^a$,
  S.~Ajimura$^b$,
  K.~Aoki$^c$,
  A.~Banu$^d$,
  H.~C.~Bhang$^e$,
  T.~Fukuda$^f$,
  O.~Hashimoto$^g$,  
  J.~I.~Hwang$^e$,
  S.~Kameoka$^g$,  
  B.~H.~Kang$^e$,
  E.~H.~Kim$^e$,
  J.~H.~Kim$^e$,
  M.~J.~Kim$^e$,
  T.~Maruta$^h$,  
  Y.~Miura$^g$,
  Y.~Miyake$^b$,
  T.~Nagae$^c$,
  M.~Nakamura$^h$,  
  S.~N.~Nakamura$^g$,
  H.~Noumi$^c$,
  Y.~Okayasu$^g$,
  H.~Outa$^i$,
  H.~Park$^j$,
  P.~K.~Saha$^f$,
  Y.~Sato$^c$,
  M.~Sekimoto$^c$,
  T.~Takahashi$^g$,
  H.~Tamura$^g$,
  K.~Tanida$^i$,
  A.~Toyoda$^c$,
  K.~Tsukada$^g$,
  T.~Watanabe$^g$,
  H.~J.~Yim$^e$ \\
  \vspace{2mm}
  $^a$Department of Physics, Tokyo Institute of Technology,
  Tokyo, 152-8551, Japan \\
  $^b$Department of Physics, Osaka University, Osaka, 560-0043, Japan \\
  $^c$High Energy Accelerator Research Organization (KEK),
  Ibaraki, 305-0801, Japan \\
  $^d$Gesellschaft f$\ddot{\mbox{u}}$r Schwerionenforschung mbH (GSI),
  Darmstadt, 64291, Germany \\
  $^e$Department of Physics, Seoul National University, Seoul, 
  151-742, Korea \\  
  $^f$Osaka Electro-Communication University, Osaka, 572-8530, Japan \\
  $^g$Department of Physics, Tohoku University, Miyagi, 980-8578, Japan \\
  $^h$Department of Physics, University of Tokyo, Tokyo, 113-0033, Japan \\
  $^i$The Institute of Physical and Chemical Research (RIKEN),
  Saitama, 351-0198, Japan \\
  $^j$Korea Reserch Institute of Standards and Science (KRISS),
  Daejeon, 305-600, Korea \\
  }
  
\begin{document}
\maketitle

\begin{abstract}
 We precisely measured \piz ~branching ratios
 of \HeLam ~and \CLam ~hypernuclei
 produced via \pik ~reaction.
 Using these $\pi^0$ branching ratios
 with the \pim ~branching ratios and the lifetimes,
 we obtained the \piz ~decay widths
 and the non-mesonic weak decay widths
 at high statistics with the accuracy of $\sim$5\%(stat)
 for both hypernuclei.
\end{abstract}


 \section{Introduction}
 

 It is well known that
 a $\Lambda$ hyperon in free space decays into a nucleon
 associated with a pion
 (\Gpim: $\Lambda \to p\pi^{-}$, \Gpiz: $\Lambda \to n \pi^{0}$).
 A bound $\Lambda$ in a nucleus ($\Lambda$ hypernucleus)
 also decays via the decay process,
 called the mesonic weak decay of $\Lambda$ hypernucleus.
 The relevant momentum transfer is only $\sim$100 MeV/$c$
 which is not necessarily enough high to exceed the Fermi momentum.
 The decay rate is therefore suppressed
 due to the Pauli blocking effect
 on the outgoing nucleon.
 Especially in light hypernuclei,
 it is sensitive to overlap
 of the $\Lambda$ wave function with the nucleus.
 Thus, the mesonic weak decay width of light hypernuclei
 gives significant information to investigate
 $\Lambda$-nucleus potential shape.

 It is believed that the folding potential
 between $\Lambda$ and light (s-shell) nuclei has a central repulsive core,
 which is calculated from the commonly used YNG $\Lambda N$ interaction.
 However, it has not been confirmed so far experimentally.
 For comparison of $\alpha$-$\Lambda$ potential,
 Motoba \et ~calculated the mesonic decay widths
 of \HeLam ~for two different potentials \cite{Mot94N}.
 One is derived from the YNG interaction,
 and the other is a simply attractive potential
 derived from one-range-gaussian two-body interaction
 called ``ORG'',
 which are determined to reproduce the $\Lambda$ binding energy of \HeLam.
 According to their calculation,
 the difference in the decay widths between them
 is $\sim$20\% as shown in Table \ref{table1}.
 However, existent experimental data cannot distinguish
 the two due to the large error.
 In the present experiment,
 we precisely measured both mesonic decay widths,
 \Gpim\ \cite{Kam03} and \Gpiz. 

 On the other hand,
 the bound $\Lambda$ in a nucleus can interact with a neighboring
 nucleon (\Gp: $\Lambda p \to np$, \Gn: $\Lambda n \to nn$),
 called the non-mesonic weak decay of $\Lambda$ hypernucleus.
 It gives unique opportunity to study baryon-baryon weak interaction
 which is hidden by strong interaction in normal nuclei.
 The total non-mesonic weak decay width (\Gnm) is one of the most
 important observables for the study.
 It is difficult to measure the \Gnm ~(= \Gp$+$\Gn) directly
 due to experimental difficulties such as final state interaction effect.
 Thus, the \Gnm\ is usually obtained by subtracting
 the mesonic weak decay widths from the total decay width \Gtot ~(inverse
 of the lifetime), as \Gnm $=$ \Gtot$-$\Gpiz$-$\Gpim.
 Table \ref{table1} shows the latest experimental data
 for \HeLam ~and \CLam.
 The errors of \Gnm\ for both hypernuclei mainly come from those
 of \Gpiz,
 so that the precise measurement of \Gpiz ~is awaited.
 
 In the present paper,
 we concentrate the measurement of $\pi^{0}$ branching ratios,
 and \Gpiz s and \Gnm s for \HeLam ~and \CLam ~are derived from
 the $\pi^-$ branching ratios and the lifetimes.

 \begin{table}[t] 

  \caption{Previous experimental results and theoretical calculations
  for the \HeLam\ and \CLam. }
  \label{table1}
  \newcommand{\m}{\hphantom{$-$}}
  \begin{tabular}{ll|lll|l}
   \hline
   \hline
   & Refs. &
   \m\Gtot / \GLam & \m\Gpim / \GLam & \m\Gpiz / \GLam & \m\Gnm  / \GLam \\
   \hline
   \HeLam\ (exp.) & \cite{Szy91}&
   \m1.03$\pm$0.08 & \m0.44$\pm$0.11 &
   \m0.18$\pm$0.20 & \m0.41$\pm$0.14 \\
   \HeLam\ (ORG) & \cite{Mot94N}&  & \m0.321 & \m0.177 &  \\
   \HeLam\ (YNG) & \cite{Mot94N}&  & \m0.393 & \m0.215 &  \\
   \hline
  \CLam\ (exp.) & \cite{Sat03}\cite{Sak91}&
   \m1.14$\pm$0.08  & \m0.113$\pm$0.015  &
   \m0.200$\pm$0.068 & \m0.828$\pm$0.087 \\
   \hline
   \hline
  \end{tabular}\\[2pt]
  \vspace*{-20pt}
 \end{table}

 \section{Experimental Method}
 
 The present experiments (E462/E508) were performed at the K6 beam line
 of the KEK 12-GeV proton synchrotron (KEK-PS).
 Hypernuclei of $^5_{\Lambda}$He and $^{12}_{\Lambda}$C were produced
 by the ($\pi^+$,$K^+$) reaction at 1.05 GeV/$c$
 on $^6$Li and $^{12}$C (active) targets.
 The hypernuclear mass spectra were calculated by
 reconstructing momenta of  incoming $\pi^+$ and outgoing $K^+$
 using the beam line spectrometer (QQDQQ)
 and the SKS spectrometer, respectively.

 The schematic view of the decay counter system is shown in Ref \cite{Out03}.
 Neutral decay particles were detected by the T4 counter arrays
 comprising 6 layers of 5$cm$-thick plastic scintillators.
 Charged decay particles were vetoed
 by thin plastic scintillators installed
 just before T4 counter arrays.
 $\pi^0$ from NMWD was identified
 by detecting high energy $\gamma$ ray,
 because the energy of this $\gamma$ ray is about 70 MeV,
 and that from other decay process is about a few MeV.
 The $\gamma$ rays were separated from neutrons by means of
 time-of-flight technique
 between the start timing counter of incident beam and the T4 counter.
 
 \section{Analysis and Results}

\begin{figure}[htbp]
 \begin{minipage}[t]{75mm}
  \includegraphics[width=1.00\linewidth]{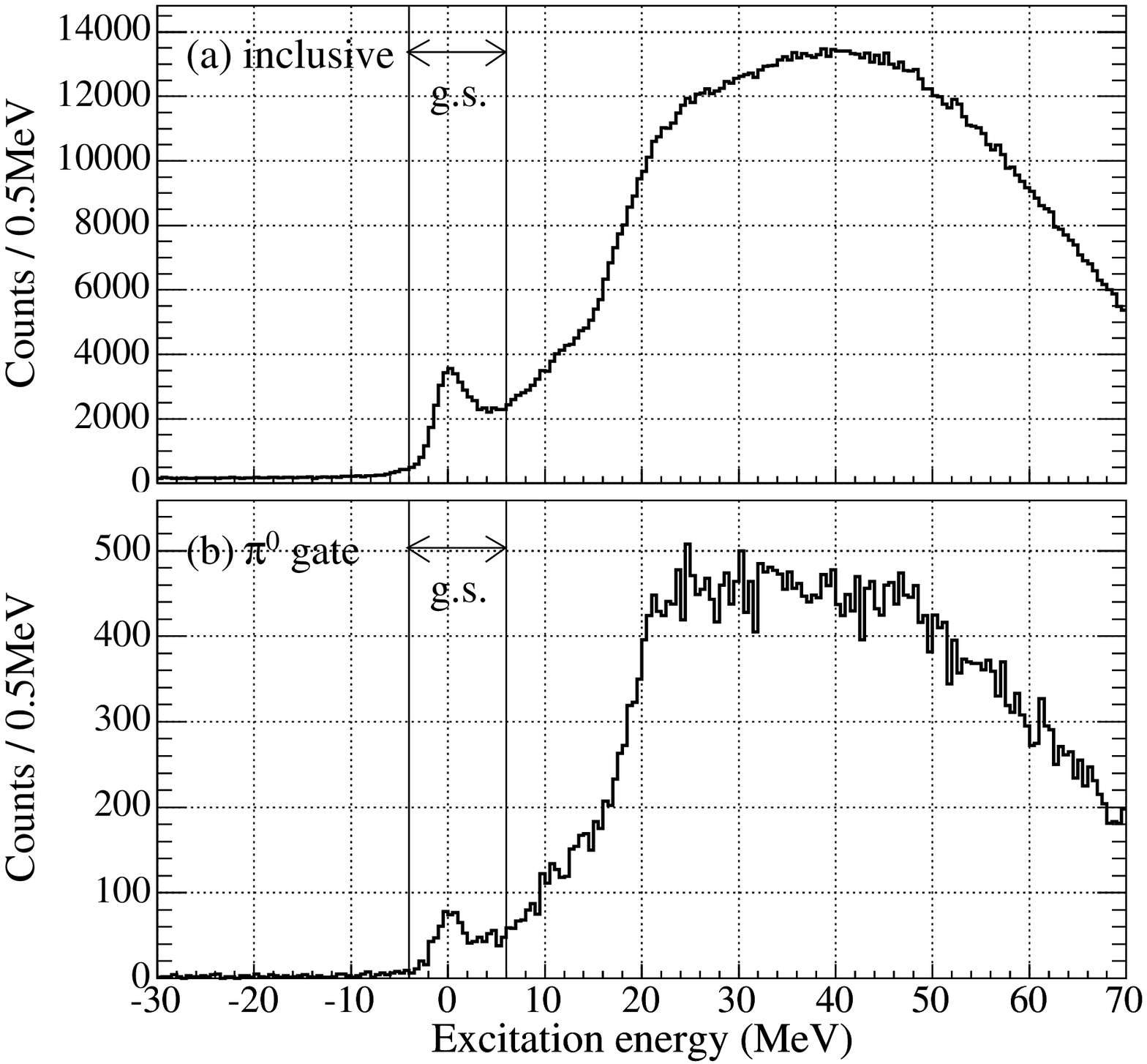}
 \end{minipage}
 \hspace{\fill}
 \begin{minipage}[t]{75mm}
  \includegraphics[width=1.00\linewidth]{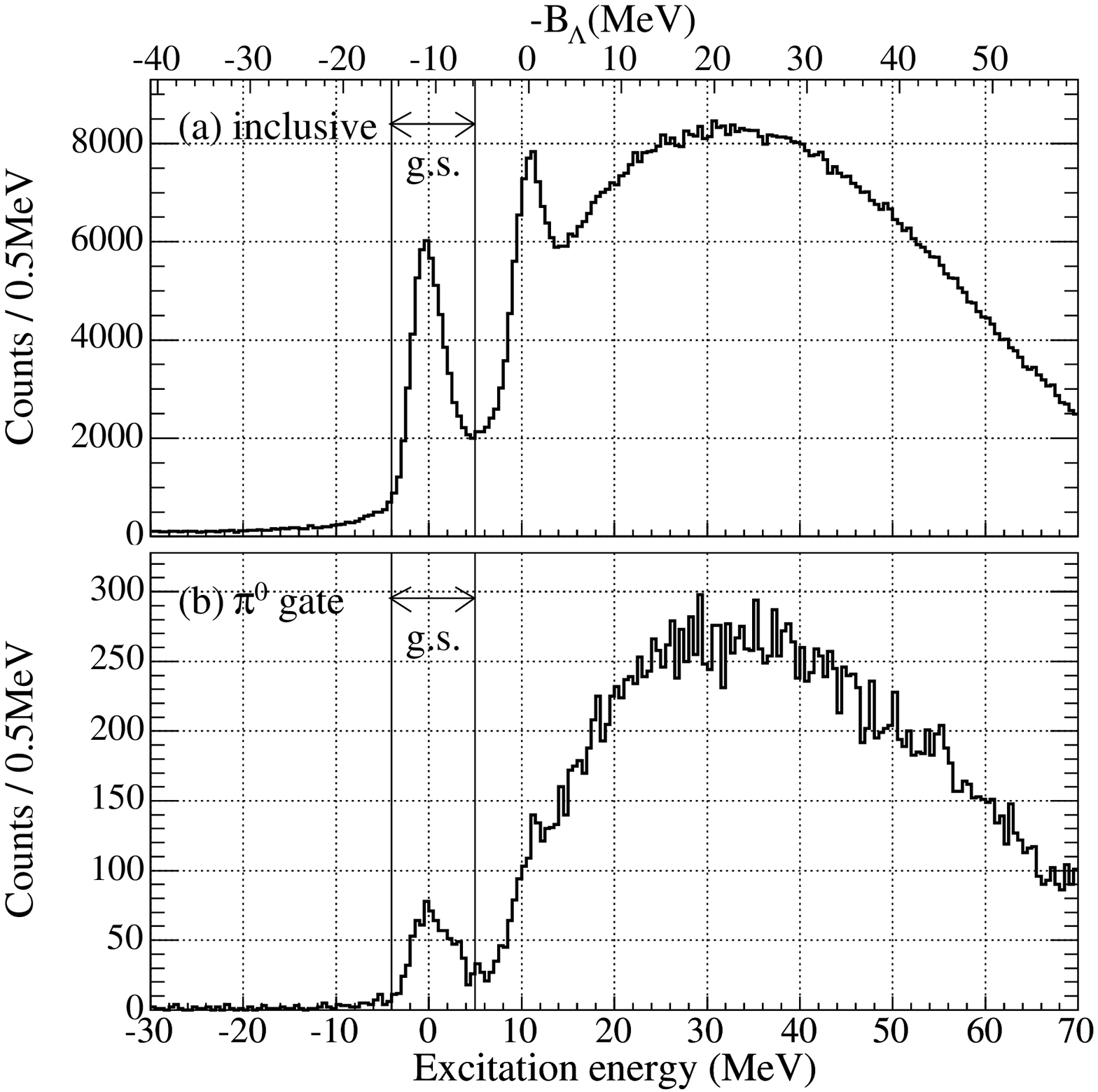}
 \end{minipage}
  \vspace*{-30pt}
 \caption{Excitation energy spectra of \LiLam ~(left figure)
 and \CLam ~(right figure).
 (a) for inclusive, (b) with the $\pi^{0}$ cut condition.}
 \label{fig:Pi0ExcitationEnergy}
\end{figure}

\begin{figure}[h]
\vspace*{-180pt}
 \begin{minipage}[t]{78mm}
  \includegraphics[width=0.95\linewidth]{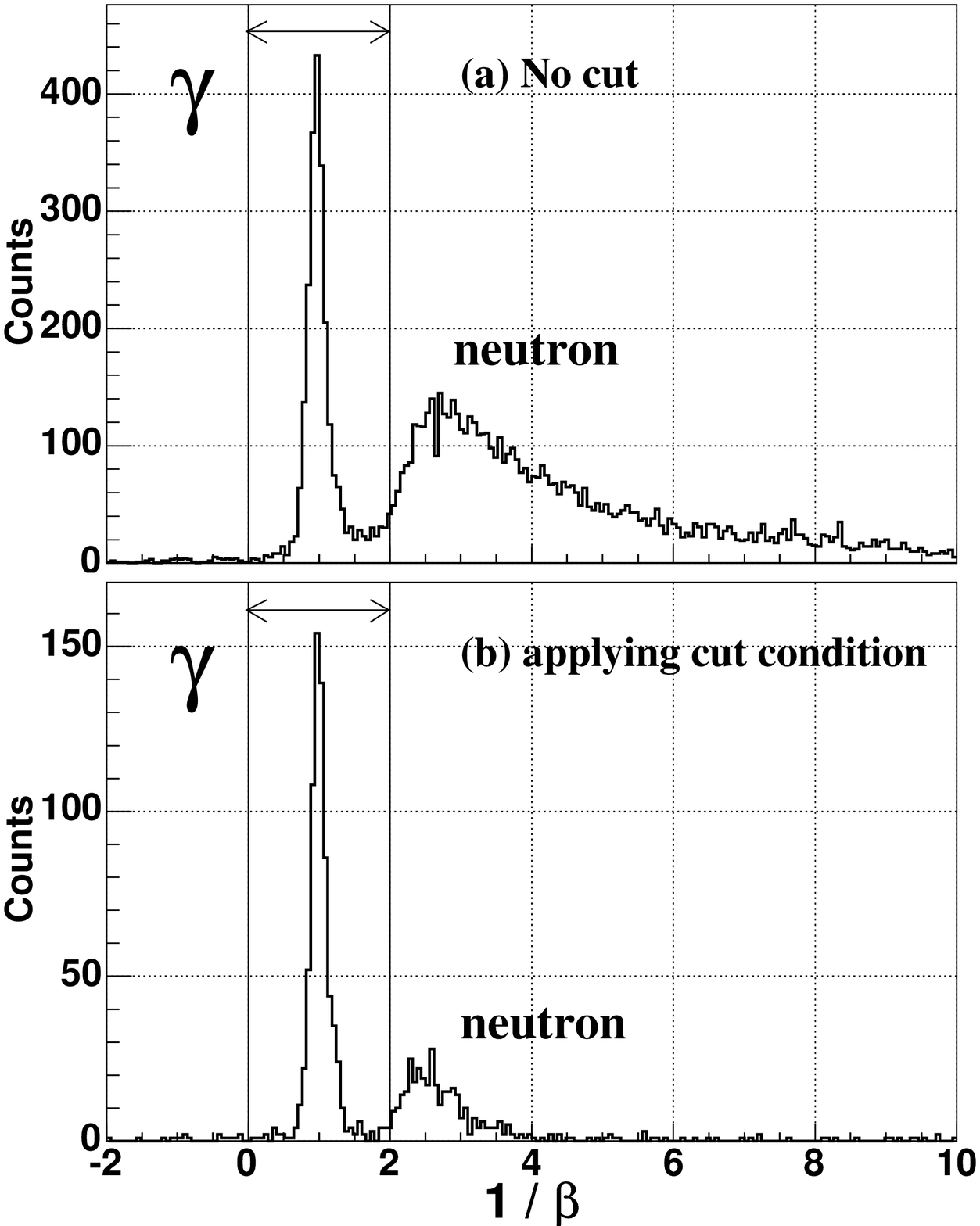}
  \vspace*{-30pt}
  \caption{1/$\beta$ spectrum of neutral particle for \CLam.
  (a) without the \piz cut condition,
  (b) with the \piz cut condition which is layer multiplicity $M \ge 2$ and 
  ADC sum $\ge$ 20 MeVee.}
  \label{fig:Pi0BetainvE508}
 \end{minipage}
 \hspace{\fill}
 \begin{minipage}[t]{78mm}
  \includegraphics[width=0.95\linewidth]{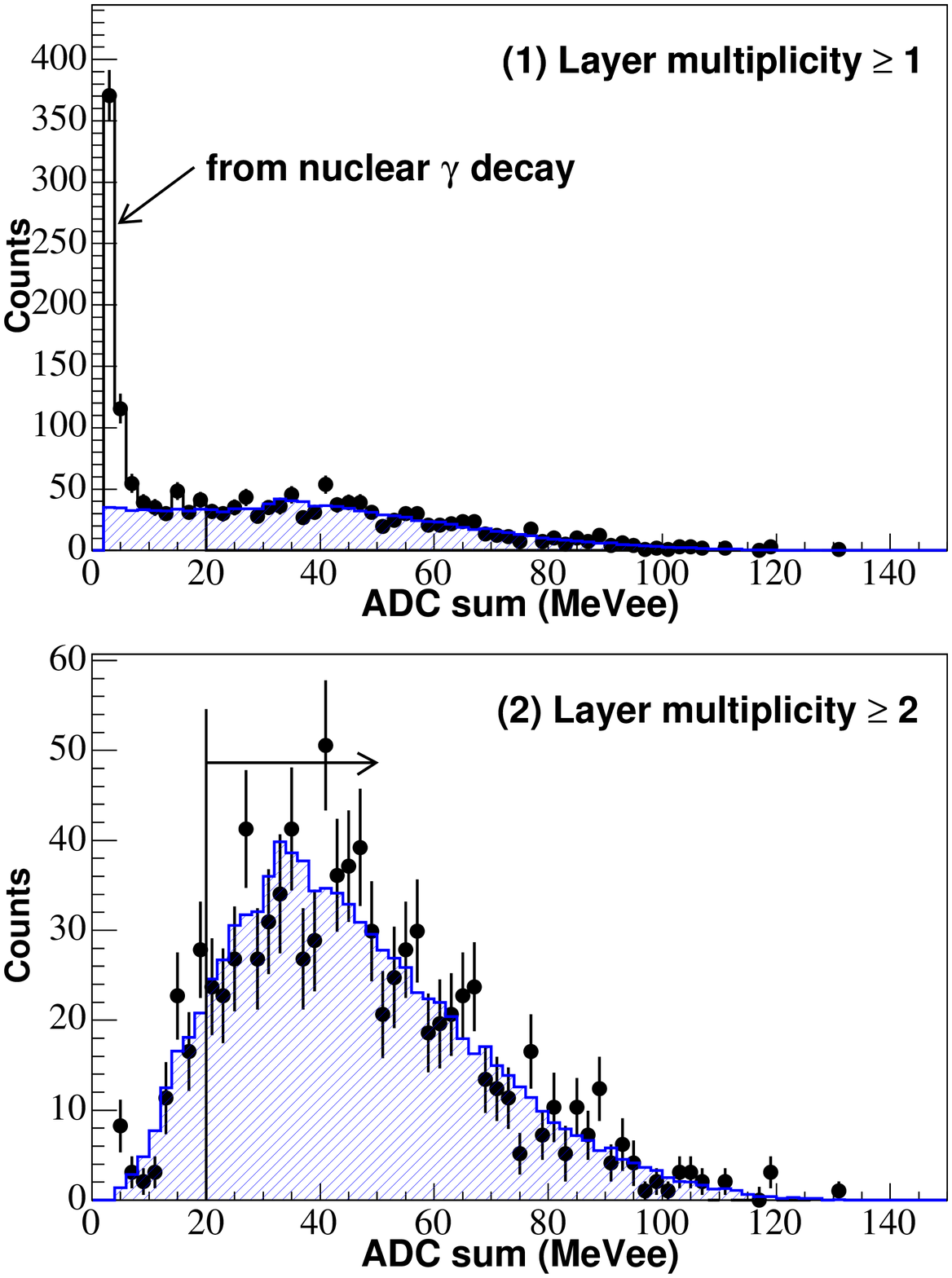}
  \vspace*{-30pt}
  \caption{$\gamma$ energy (ADC sum) spectra from \piz ~decay of
  \HeLam ~(point with error bar) are compared
  with the simulation (shaded one).
  (1) layer multiplicity $M \ge 1$, (2) $M \ge 2$.}
  \label{fig:Pi0ADCsumE462}
 \end{minipage}
\end{figure}

 The formations of each hypernucleus were identified
 by gating the ground state region
 in the excitation energy spectra of \LiLam ~and \CLam ~as shown
 in Figure \ref{fig:Pi0ExcitationEnergy}  (a).
 Neutral particles from the decay were detected at T4 counter
 with 2 MeVee (MeV electron equivalent) threshold.
 The 1/$\beta$ spectrum for $^{12}_{\Lambda}$C
 is shown in Figure \ref{fig:Pi0BetainvE508} (a),
 which shows good $\gamma$/$n$ separation.
 The $\gamma$ gate corresponds to 0 $\le$ 1/$\beta \le$ 2.
 Using the yields below the $\gamma$ peak (1/$\beta <$ 0),
 the accidental background within the $\gamma$ gate
 was estimated as good as $\sim$ 2 \%.

 In order to estimate the efficiency of the detector setup,
 the GEANT-based Monte Carlo simulation was performed.
 The efficiency depends on the energy of $\pi^0$.
 We assumed mono-energetic (104.9 MeV/$c$) $\pi^0$
 for \HeLam ~(\HeLam ~$\to \pi^0 + ^5$He (g.s.)),
 and we used the $\pi^0$ distribution for \CLam ~given by 
 Motoba \et \cite{Mot94P}.
 Figure \ref{fig:Pi0ADCsumE462} shows
 the $\gamma$ energy spectra for \HeLam.
 The points with error bars
 are the experimental data,
 and the shaded one is the simulation.
 To select $\gamma$-ray shower clearly,
 we applied the multiplicity cut for the identification.
 Upper figure shows the spectrum with applying multiplicity $M \ge 1$,
 and lower figure shows that for $M \ge 2$.
 There is a low energy background
 in the spectrum for the $M \ge 1$ condition,
 whereas the background disappear in that for $M \ge 2$ condition.
 In order to remove the low energy background completely,
 we determined the \piz\ cut condition as ``$M \ge 2$'' and
 ``ADC sum $\ge 20$ MeVee''.
 The 1/$\beta$ spectrum applied this \piz ~cut condition 
 is shown in Figure \ref{fig:Pi0BetainvE508} (b). 
 In this figure, the $\gamma$ ray from $\pi^0$ decay more clearly
 separated from neutron.
 For the efficiency estimation,
 the same cut condition was applied in the simulation.
 The detection efficiency (including the detector acceptance)
 is estimated to be $\varepsilon \sim$ 10.5 \%.
 The good agreement of the energy spectra between data and
 simulation in Figure \ref{fig:Pi0ADCsumE462} (2) shows
 that we can count for the efficiency estimation.

 Figure \ref{fig:Pi0ExcitationEnergy} (b) shows the
 excitation energy of \LiLam ~and \CLam ~with
 the \piz ~cut condition.
 The \piz ~branching ratio is represented by
 $b_{\pi^{0}} = N_{\pi^{0}} / N_{inc} / \varepsilon$,
 where $N_{inc}$ and $N_{\pi^{0}}$ are 
 the numbers gated for the ground state regions
 shown in the Figure \ref{fig:Pi0ExcitationEnergy} (a) and (b), respectively.
 Consequently, the \piz ~branching ratios for \HeLam ~and \CLam ~were
 determined to be
 $b_{\pi^-}$ = 0.212 $\pm$ 0.008 and 0.133 $\pm$ 0.005,
 respectively (statistical error only), though preliminary yet.

 \vspace*{10pt}

 \Gpiz s and \Gnm s for \HeLam ~and \CLam ~were derived
 from our results of the lifetimes and
 the $\pi^{-}$ branching ratios\cite{Kam03}
 as shown in Table \ref{table2}.
 The result of \Gpiz ~for \HeLam ~is located in
 between those of ORG- and YNG-based calculations,
 which is consistent with the \Gpim ~result \cite{Kam03}.
 It indicates that the $\alpha$-$\Lambda$ overlapping
 is larger than that of the YNG-based calculation.
 
 The statistical errors of obtained \Gnm s for \HeLam ~and \CLam ~were
 much improved as 
 34\% $\to$ 5\% for \HeLam\ and
 11\% $\to$ 5\% for \CLam.
 The theoretical calculations of non-mesonic weak decay
 are required to meet these $\Gamma_{nm}$ results and 
 our $\Gamma_n/\Gamma_p$ results\cite{Out03} simultaneously.
 
 \begin{table}[htb]
  \vspace*{-0.5cm}
  \caption{Summary of present preliminary results for the \HeLam\ and \CLam.}
  \label{table2}
  \newcommand{\m}{\hphantom{$-$}}
  \begin{tabular}{l|lll|l}
   \hline
   \hline
   & 
   \m\Gtot / \GLam & \m\Gpim / \GLam & \m\Gpiz / \GLam & \m\Gnm  / \GLam \\
   \hline
   \HeLam\ &
   0.947$\pm$0.038 \cite{Kam03} & 0.340$\pm$0.016 \cite{Kam03} &
   {\bf 0.201$\pm$0.011}        & {\bf 0.406$\pm$0.020} \\

  \CLam\ & 
   1.242$\pm$0.042 \cite{Kam03}  & 0.123$\pm$0.015 \cite{Sat03}\cite{Kam03} &
   {\bf 0.165$\pm$0.008} & {\bf 0.953$\pm$0.032} \\

   \hline
   \hline
  \end{tabular}\\[2pt]
 \end{table}

\vspace*{150pt}


\end{document}